\documentclass{ws-p9-75x6-50}

\begin{document}

\title{Quantum Monte Carlo
Methods for Nuclei at Finite Temperature}

\author{Y. Alhassid}

\address{Center for Theoretical Physics, Sloane Physics
Laboratory,
Yale University, New Haven, Connecticut  06520,
USA\\E-mail: yoram.alhassid@yale.edu}


\maketitle

\abstracts{We discuss finite temperature quantum Monte Carlo methods in the 
framework  of the interacting nuclear shell model.  The methods are based on a 
representation  of the imaginary-time many-body propagator as a  superposition 
of  one-body propagators describing non-interacting fermions moving in 
fluctuating  auxiliary fields. Fermionic Monte Carlo calculations have been 
limited  by a ``sign'' problem.  A practical solution in the nuclear case 
enables  realistic calculations in much larger configuration spaces than can 
be  solved by conventional methods.  Good-sign interactions can be constructed 
for  realistic estimates of certain nuclear properties. We present various 
applications  of the methods for calculating collective properties and level 
densities. 
}

\section{Introduction}

   A variety of models have been developed over the years to explain the 
observed   properties of  nuclei in various mass regions.
One of the more fundamental of these models is the interacting shell model,
where valence nucleons  (outside closed shells) move in a mean field 
potential  and interact via a residual nuclear force. This effective 
interaction  can be traced back to the bare nucleon-nucleon interaction via 
Brueckner's  $G$-matrix.

The nuclear shell model was applied
successfully in the description of $p$\cite{CK65}, $sd$\cite{BW}, and lower 
$pf$-shell  nuclei\cite{RB91,CPZ,Akiva}.
However, the size of the model space increases combinatorially with the  
number  of valence nucleons and/or orbitals, and conventional diagonalization 
of  the nuclear
Hamiltonian in a full shell is limited to nuclei up to
 $A \sim 50$.

 At finite temperature many levels contribute, and diagonalization methods 
are  even more limited. However, there are methods that do not require direct 
diagonalization  of the Hamiltonian.  At  temperature $T$ the nucleus is 
described  by its free energy $F(T) = - T \ln Z(T)$,
where $Z(T)  =  {\rm Tr} e^{ -H/T}$ is the partition function expressed in 
terms  of the nuclear Hamiltonian $H$.
It is difficult to calculate the partition function because $H$ includes 
interactions  that are strong, and non-perturbative methods are required. 
Thermal  mean-field approximations, e.g., the Hartree-Fock approximation, are 
useful  and tractable. Small amplitude fluctuations around the mean field can 
also  be taken into account in the random phase approximation (RPA).  However, 
in  the finite nuclear system, large amplitude fluctuations are important. In 
fact,  interaction effects can be taken into account exactly when all 
fluctuations  of the mean field are properly included. This is formally 
expressed  by the Hubbard-Stratonovich transformation.\cite{HS57} In this 
transformation  the nuclear  propagator in imaginary time is represented
as a superposition of one-body propagators  that describe
 non-interacting fermions moving in fluctuating auxiliary fields.  Quantum 
Monte  Carlo methods were developed to carry out the integration over the 
large  number of auxiliary  fields.\cite{LJK93,ADK94}
These calculations are computationally intensive and became feasible only 
with  the introduction of parallel computers. Similar methods
were developed for strongly correlated electron systems.\cite{LG92,Linden}

The initial applications  of the Monte Carlo techniques
were severely limited by the ``sign'' problem, which is generic to all 
fermionic  Monte Carlo methods and occurs for all  realistic nuclear 
interactions.  We have
developed a practical solution to this sign problem\cite{ADK94}
that made possible realistic calculations in much larger configuration spaces 
than  could be
treated previously.

Other quantum Monte Carlo methods were developed for the many-nucleon system. 
Variational  and Green function Monte Carlo methods were used successfully to 
derive  properties of light nuclei from the bare nucleon-nucleon 
force.\cite{PPC}  Recently, a hybrid Monte Carlo method was suggested, in 
which  the spin variables are decoupled using the Hubbard-Stratonovich 
transformation.\cite{SF99} 

  The shell model Monte Carlo (SMMC) methods are reviewed in Section 
\ref{methods}.  Some of their applications are presented in Sections 
\ref{Gamow-Teller},  \ref{soft} and \ref{densities}.

\section{Methods}\label{methods}

\subsection{The Hubbard-Stratonovich transformation}\label{subsec:HS}

 The shell model Hamiltonian contains a one-body part and a residual two-body 
interaction.  A generic Hamiltonian with two-body interactions can be brought 
to  the form
\begin{eqnarray}\label{H}
 H = \sum_a \epsilon_a \hat{n}_a
 +  \frac{1}{2} \sum_\alpha {v_\alpha} \hat{\rho}_\alpha^2 \; ,
 \end{eqnarray}
where $\epsilon_a $ is the single-particle energy in orbital $a$,
$\hat{\rho}_\alpha$ are linear combinations of  one-body densities
 $\hat{\rho}_{ij} = a_i^\dagger a_j$,  and  $v_\alpha $ are interaction
``eigenvalues.''

 The canonical density operator  $e^{-\beta H}$, where $\beta=1/kT$ is the 
inverse  temperature, can also be viewed as
 the many-body evolution operator in
imaginary time $\beta$.  The Hubbard-Stratonovich (HS) 
transformation\cite{HS57}  is an exact representation
 of this propagator as a path integral over one-body propagators in
 fluctuating external fields $\sigma(\tau)$.
It is derived by dividing the time interval $(0,\beta)$ into $N_t$ time 
slices  of length $\Delta \beta$ each, and rewriting
$e^{-\beta H}  =  \left(e^{-\Delta \beta H}\right)^{N_t}$.  The propagator 
for  each time slice can be written as  $e^{-\Delta \beta H} \approx  \prod_a 
e^{-  \Delta \beta \epsilon_a \hat{n}_a}
\prod_\alpha e^{-\frac{1}{2}{ \Delta}\beta {v_\alpha}
 \hat{\rho}_\alpha^2}$ to order  $({\Delta}\beta)^2$.
Each factor with  $\hat{\rho}_\alpha^2$  can be written as an integral over 
an  auxiliary variable $\sigma_\alpha$
\begin{eqnarray}
 e^{-\frac{1}{2} { \Delta}\beta {v_\alpha} \hat{\rho}_\alpha^2}
 = \sqrt{{\beta| v_\alpha|}\over{2\pi}} \int  d\sigma_\alpha
 e^{-\frac{1}{2} {\Delta}\beta |v_\alpha| {\sigma_\alpha^2}}
 e^{-{ \Delta}\beta |v_\alpha|  s_\alpha \sigma_\alpha \hat{\rho}_\alpha}
\;,
\end{eqnarray}
where $s_\alpha=\pm1$ for $v_\alpha < 0$ and $s_\alpha=\pm i$ for $v_\alpha > 
0$. 
By introducing a different set  of fields   $\sigma_\alpha(\tau_n)$ at each 
time 
slice $\tau_n = n \Delta \beta$, we obtain the HS representation of the 
propagator 
\begin{eqnarray}\label{HS}
 e^{-\beta H} = \int {\cal D}[\sigma] G_\sigma U_\sigma \;,
\end{eqnarray}
where
\begin{equation}
G_\sigma =  e^{-\frac{1}{2}{ \Delta}\beta \sum_{\alpha,n}
|v_\alpha| \sigma_\alpha^2(\tau_n)}
\end{equation}
 is a Gaussian weight, and
\begin{equation}\label{prop}
 U_\sigma =e^{-{ \Delta}\beta
 h_\sigma(\tau_n)} \ldots e^{-{ \Delta}\beta h_\sigma(\tau_1)} =
T  \exp{\left( -\int_0^\beta h_\sigma(\tau) d\tau \right)}
\end{equation}
is the propagator of non-interacting nucleons moving in time-dependent 
external 
 one-body auxiliary fields $\sigma_\alpha(\tau)$. The one-body Hamiltonian in
 (\ref{prop}) is given by
\begin{eqnarray}
h_\sigma(\tau) = \sum_a \epsilon_a \hat{n}_a
 +\sum_\alpha s_\alpha |v_\alpha| \sigma_\alpha(\tau) \hat{\rho}_\alpha  \;,
\end{eqnarray}
 and the metric in the functional integral (\ref{HS}) is 
\begin{eqnarray}
 {\cal D}[\sigma] \equiv \Pi_{\alpha, n}\left[d\sigma_\alpha
(\tau_n)  \sqrt{{{\Delta}\beta |v_\alpha| /{2\pi}}}\right]
\;.
\end{eqnarray}

The thermal expectation
 of an observable  $O$ can be represented from the HS transformation 
(\ref{HS})  as
\begin{eqnarray}\label{observ}
\langle O \rangle=
{{\rm Tr}\,( O e^{-\beta H})\over{\rm Tr}\,(e^{-\beta H})}=
{\int {\cal D}[\sigma] G_\sigma \langle O \rangle_\sigma{\rm Tr}\,U_\sigma
\over \int {\cal D}[\sigma] G_\sigma {\rm Tr}\,U_\sigma}  \;,
\end{eqnarray}
 where  $\langle O \rangle_\sigma\equiv
 {\rm Tr} \,( O U_\sigma)/ {\rm Tr}\,U_\sigma$ is the expectation value of
$O$ for non-interacting particles moving in external fields $\sigma(\tau)$.

 The calculation of the integrands in (\ref{observ}) can be reduced to matrix 
algebra  in the single-particle space. To show that, we denote by $N_s$ the 
number  of valence single-particle orbits, and represent the one-body 
propagator  $U_\sigma$ in the single-particle space by an $N_s \times N_s$ 
matrix  ${\bf U}_\sigma$.  The grand-canonical trace of $U_\sigma$ in the 
many-particle  fermionic space can be expressed directly in terms of ${\bf 
U}_\sigma$ 
\begin{equation}\label{partition}
{\rm Tr}\; U_\sigma = \det ( {\bf 1} + {\bf U}_\sigma) \;.
\end{equation}
Eq. (\ref{partition}) is just the grand-canonical partition function of 
non-interacting  fermions (moving in time-dependent external fields $\sigma$).

Similarly the grand-canonical expectation value of a one-body operator $O = 
\sum_{i,j}  O_{ij} a^\dagger_i a_j$ can be calculated from
\begin{equation}\label{1-body}
\langle a_i^\dagger a_j \rangle_\sigma
 = \left[ {1 \over {\bf 1} +{\bf U}^{-1}_\sigma }
\right]_{ji} \;.
\end{equation}
The grand-canonical expectation value of a two-body operator can be 
calculated  using Wick's theorem
$\langle a^\dagger_i a^\dagger_j a_l a_k \rangle = \langle a^\dagger_i a_l 
\rangle  \langle
 a^\dagger_j a_k\rangle - \langle a^\dagger_i a_k \rangle  \langle
 a^\dagger_j a_l\rangle$ together with (\ref{1-body}).

 In practice, we are interested in canonical expectation values, for which 
the  number of particles ${\cal A}$ is fixed.  Canonical quantities can be 
calculated  by an exact particle-number projection. The canonical (one-body) 
partition  function is given as a Fourier transform\cite{Ormand94}
\begin{eqnarray}\label{canonical}
{\rm Tr}_{\cal A} U_\sigma =\frac{e^{-\beta\mu {\cal A}}
}{N_s}\sum_{m=1}^{N_s}
e^{-i\phi_m {\cal A}}\det \left[ {\bf 1}+e^{i\phi_m}e^{\beta\mu}{\bf 
U}_\sigma\right] 
\;,
\end{eqnarray}
where $\phi_m=2\pi m/N_s \;\; (m=0,\ldots,N_s)$ are quadrature points and 
$\mu$ 
is a chemical potential.  Similarly for a one-body observable $O$
\begin{equation}
{\rm Tr}_{\cal A} \left(O U_\sigma \right) =\frac{e^{-\beta \mu {\cal 
A}}}{N_s}  \sum_{m=1}^{N_s}
e^{-i\phi_m{\cal A}}
\det \left[ {\bf 1}+e^{i\phi_m +\beta\mu}{\bf U}_\sigma\right] {\rm 
tr}\;\left(  {1 \over {\bf 1} +e^{-i\phi_m-\beta\mu}{\bf U}^{-1}_\sigma } {\bf 
O} 
\right) \;,
\end{equation}
where ${\bf O}$ is the matrix with elements $O_{ij}$.
In practical calculations it is necessary to project on both neutron number 
$N$  and proton number $Z$, and in the following ${\cal A}$ will denote 
$(N,Z)$. 

\subsection{Monte Carlo methods}\label{MC}

  The integrands in (\ref{observ}) are easily calculated by matrix algebra in 
the  single-particle space (see, e.g., Eq. (\ref{partition}) and 
(\ref{1-body})). 
However, the number of integration variables $\sigma_\alpha (\tau)$ is very 
large. 
 For small but finite $\Delta \beta$,  this multi-dimensional integral can be 
evaluated  exactly (up to a statistical error) by Monte Carlo 
methods.\cite{LJK93} 

The first step is to define a positive-definite weight function
\begin{equation}
 W_\sigma \equiv
  G_\sigma \vert {\rm Tr} \; U_\sigma \vert \;.
\end{equation}
 We can rewrite the expectation value (\ref{observ}) of an observable $O$ as
\begin{equation}\label{observ-W}
\langle O\rangle={\int D[\sigma]W_\sigma\Phi_\sigma
\langle O\rangle_\sigma\over
\int D[\sigma]W_\sigma\Phi_\sigma}  \;,
\end{equation}
where
\begin{equation}\label{sign}
\Phi_\sigma\equiv {\rm Tr} \; U_\sigma /\vert {\rm Tr}\; U_\sigma \vert
\end{equation}
 is the sign of the one-body partition function.

   In the Monte Carlo approach, a random walk is performed in the space of 
auxiliary  fields $\sigma \equiv\{\sigma_\alpha(\tau_m)\}$
 that samples the $\sigma$-fields
 according to  the distribution $W_\sigma$.  The $W$-weighted average of any 
$\sigma$-dependent  quantity $X_\sigma$  can be estimated from
\begin{equation}\label{X-average}
\langle X_\sigma \rangle_W \equiv
{\int D[\sigma]W_\sigma X_\sigma
\over
\int D[\sigma]W_\sigma} \approx  {1\over M} \sum_k
X_{\sigma^{(k)}} \;,
\end{equation}
where $\sigma^{(k)}$ are $M$ uncorrelated samples.
The statistical error of $\langle X_\sigma \rangle_W$ can be estimated from 
the  variance of the ``measurements'' $X_{\sigma^{(k)}}$. Though a standard 
random  walk can be constructed by  the Metropolis algorithm,
a modification based on Gaussian quadratures
 improves the efficiency in the nuclear case.\cite{Dean93}

  Using (\ref{observ-W}) and (\ref{X-average}), thermal expectation values 
are  calculated from
\begin{equation}\label{average}
\langle O \rangle ={ \left\langle \langle O \rangle_\sigma \Phi_\sigma 
\right\rangle_W  \over \langle \Phi_\sigma \rangle_W}
\approx   {\sum_k \langle O \rangle_\sigma  \Phi_\sigma  \over \sum_k 
\Phi_\sigma  } \;.
\end{equation}

Often  the sign $\Phi_\sigma$ in (\ref{sign}) is not positive for some 
samples  $\sigma$,
  and we must  ``measure'' it
 together with the observable $\langle O\rangle_\sigma$.

\subsection{The Monte Carlo sign problem}

  When the average sign $\langle\Phi_\sigma\rangle_W$ is
smaller than its uncertainty, the Monte Carlo method fails. This leads to the 
so-called  Monte Carlo ``sign'' problem:
 the sign of the integrand fluctuates among samples,  and the integral is
the result of a delicate cancellation that cannot be reproduced with a finite 
number  of samples. Often the average sign decreases exponentially with 
$\beta$,  and the problem becomes more severe at low temperatures.  The sign 
problem   is generic to  all fermionic Monte Carlo methods\cite{LG92} and  has 
severely  limited their applications.
 In particular, the problem occurs for all realistic
effective interactions in the nuclear shell model.  We have developed a 
practical  solution to the sign problem in the nuclear case that allows 
realistic  calculations in very large model spaces.\cite{ADK94}

Certain interactions can be shown to have a ``good'' sign. A  time-reversal 
invariant  Hamiltonian  with two-body interactions can be
 written as
\begin{eqnarray}\label{trev}
H=\sum_a \epsilon_a \hat{n}_a+
{1\over2}\sum_\alpha v_\alpha
\left({\rho}_\alpha \bar{\rho}_\alpha  +  \bar{\rho}_\alpha
{\rho}_\alpha\right)\;,
\end{eqnarray}
where $\bar{\rho}_\alpha$ is the time-reverse of ${\rho}_\alpha$ and
 $v_\alpha $ are real.  When all  $v_\alpha$ are
{\em negative}
in the representation (\ref{trev}), the grand-canonical partition function is 
positive-definite  for any sample $\sigma$
 and the interaction has  a good sign. The proof is based on a
generalization of Kramer's degeneracy to the complex plane.
The one-body Hamiltonian that appears in the HS decomposition for 
(\ref{trev})  has the form
\begin{equation}\label{1body-t}
h_\sigma =\sum_a  \epsilon_a \hat{n}_a  +
\sum_\alpha \left(v_\alpha s_\alpha {\rho}_\alpha  +
 v_\alpha s_\alpha \bar{\rho}_\alpha \right) \;.
\end{equation}
 When all $v_\alpha < 0$,  $s_\alpha=1$, and  the Hamiltonian (\ref{1body-t}) 
is  time-reversal invariant  $\bar h_\sigma = h_\sigma$. The
 eigenstates of $\bf{U}_\sigma$
 appear then in time-reversed pairs with complex conjugate
 eigenvalues $\{\lambda_i,\lambda^\ast_i\}$, and
the  grand-canonical partition function  ${\rm Tr}\;U_\sigma=\det (1 + {\bf 
U}_\sigma)=\Pi_i  |1 +\lambda_i|^2$ is positive-definite.
 The canonical partition function for
 even-even nuclei is also positive-definite.

Realistic effective nuclear interactions have both positive and negative 
``eigenvalues''  $v_\alpha$, leading to a severe sign problem at low $T$. 
However,  the eigenvalues that are large in magnitude are all negative and 
therefore  have a good sign.  This property of realistic nuclear forces can be 
used  in several ways to overcome the sign problem.

In general, the Hamiltonian is decomposed into ``good'' and ``bad'' parts: 
$H=H_G+H_B$.  The good Hamiltonian contains the one-body part plus the 
two-body  terms with $v_\alpha <0$, while the bad Hamiltonian contains the 
two-body  terms with $v_\alpha >0$.  We construct a family of Hamiltonians
\begin{equation}\label{g-Hamiltonian}
H_g=H_G+g H_B
\end{equation}
that depends on a continuous coupling parameter $g$. For any $g <0$, the 
Hamiltonian  (\ref{g-Hamiltonian}) is good-sign and accurate Monte Carlo 
calculations  can be performed for    $\langle O \rangle_g  \equiv {\rm 
Tr}\,(O  e^{-\beta
H_g})/{\rm Tr}\,(e^{-\beta H_g})$.  The dependence of $\langle O \rangle_g$  
on  $g$ is expected to be ``smooth'' and can therefore be extrapolated to 
$g=1$.   We found, through tests in the $sd$- and lower $pf$-shell, that 
low-degree  polynomial extrapolations (usually first or second order) are 
sufficient.  The extrapolation technique is used in Section 
\ref{Gamow-Teller}. 

Certain nuclear properties such
 as collectivity and level densities can be reliably calculated by 
constructing  nuclear Hamiltonians that are completely free of the sign 
problem.\cite{ABDK,NA97}  These Hamiltonians include correctly the dominating 
collective  components of realistic effective interactions.\cite{DZ96}
An example of a good-sign Hamiltonian, used in the applications discussed in 
Sections  \ref{soft} and \ref{densities}, is
\begin{eqnarray}\label{Hamiltonian}
  H = \sum_a \epsilon_a \hat n_a - g_0 P^{(0,1)\dagger}\cdot \tilde P^{(0,1)}
     - \chi \sum_\lambda k_\lambda O^{(\lambda,0)}\cdot O^{(\lambda,0)} \;,
\end{eqnarray}
where
\begin{eqnarray}
 P^{(\lambda,T)\dagger}&=&{\sqrt{4\pi}\over{2(2\lambda +1)}}
 \sum_{ab} \langle j_a\| Y_\lambda \|j_b\rangle
 [a_{j_a}^\dagger \times a_{j_b}^\dagger]^{(\lambda,T)}\;, \nonumber\\
O^{(\lambda,T)}&=&{1\over\sqrt{2\lambda +1}}
 \sum_{ab} \langle j_a\| {{dV}\over{dr}} Y_\lambda \|j_b\rangle
[a_{j_a}^\dagger \times \tilde a_{j_b}]^{(\lambda,T)} \;.
\end{eqnarray}
The modified annihilation operator is defined
by $\tilde a_{j,m,m_t} = (-)^{j-m+{1\over 2}-m_t} a_{j,-m,-m_t}$,
and a similar definition is used for $\tilde P^{(\lambda,T)}$.
The single-particle energies $\epsilon_a$ are calculated from a Woods-Saxon 
potential  plus spin-orbit interaction.\cite{ref:BM}
$V$ in (\ref{Hamiltonian}) is the central part of this  potential, and the 
multipole  interaction is obtained
by expanding the  surface-peaked interaction
$v({\bf r}, {\bf r}^\prime)
 = -\chi (dV/dr)(dV/dr^\prime) \delta(\hat{\bf r} - \hat{\bf r}^\prime)$.
 Since the monopole pairing and isoscalar multipole-multipole interactions 
are  all attractive, they lead to a good-sign Hamiltonian (\ref{Hamiltonian}).

Experimental gap parameters (found from  odd-even mass differences) are used
 to determine the pairing strength $g_0$ through a particle-projected BCS 
calculation.  The parameter $\chi$ of the
surface-peak interaction is determined self-consistently:
$\chi^{-1} = \int_0^\infty dr \; r^2  \left(dV/ dr \right)  \left(d\rho/ dr 
\right)$,  where $\rho$ is the nuclear density. The renormalization factors 
$k_\lambda$  account for core polarization.

\subsection{Particle-number reprojection}\label{reproject}

The SMMC methods are computationally intensive. Calculations can be done more 
efficiently  in the particle-reprojection technique.\cite{ALN99}
Suppose we use Monte Carlo methods to sample a nucleus with ${\cal A}$ 
particles.   The ratio $Z_{{\cal A}^\prime}/Z_{\cal A}$
between  the partition function of a nucleus with
${\cal A}'$ particles and the partition function of the original
nucleus can be calculated from
\begin{eqnarray}\label{partition-ratio}
{Z_{{\cal A}^\prime}(\beta) \over Z_{\cal A}(\beta)} \equiv
{{\rm Tr}_{{\cal A}^\prime} e^{-\beta H} \over {\rm Tr}_{\cal A} e^{-\beta
H}}  =
\left\langle { {\rm Tr}_{{\cal A}^\prime} U_\sigma \over
{\rm Tr}_{\cal A} U_\sigma} \right\rangle_W \;,
\end{eqnarray}
where $\langle \ldots\rangle_W$ is defined in (\ref{X-average}).  Each of the 
partition  functions within the brackets in
(\ref{partition-ratio}) is calculated from (\ref{partition}).
The expectation value of an observable $O$ for the nucleus with
${\cal
A}'$  particles is calculated using
\begin{equation}\label{observable'}
\langle O \rangle_{{\cal A}'}= {\left\langle \left({{\rm Tr}_{{\cal A}'}
OU_\sigma  \over {\rm Tr}_{{\cal A}'} U_\sigma}\right) \left({ {\rm
Tr}_{{\cal
A}^\prime}  U_\sigma \over {\rm Tr}_{\cal A} U_\sigma}\right) \right\rangle_W
\over
\left\langle { {\rm Tr}_{{\cal A}'} U_\sigma  \over
{\rm Tr}_{\cal A} U_\sigma} \right\rangle_W
} \;.
\end{equation}
In the particle-reprojection method, the Monte Carlo sampling is carried out 
by  projecting on a nucleus with fixed ${\cal A}$, and the partition functions 
and  observables are calculated for a family of nuclei with  ${\cal A}' \neq 
{\cal  A}$ using Eqs.  (\ref{partition-ratio}) and (\ref{observable'}).

In the following Sections we present some of the applications of the shell 
model  Monte Carlo approach.

\section{Quenching of the Gamow-Teller strength}\label{Gamow-Teller}

Gamow-Teller rates, especially in the iron region, are important inputs in 
models  of  stellar collapse and  supernova formation.\cite{Bethe90} They 
determine  the electron-capture rates, which in turn affect the 
electron-to-baryon  ratio.

 The Gamow-Teller transition operator is given by
\begin{equation}
 GT^{\pm} \propto \vec\sigma  \tau_{\pm} \;,
\end{equation}
where $\vec\sigma$ and $\vec\tau$ are the spin and isospin operators, 
respectively.  The $GT^+$ transition operator changes a proton into a neutron, 
and  its strength function can be measured in $(n,p)$ reactions.\cite{GT-exp} 
The  total strength $B(GT^+)$ is found experimentally to be strongly quenched 
relative  to estimates based on the independent-particle model.

\begin{figure}[h]
\epsfxsize= 8 cm
\centerline{\epsfbox{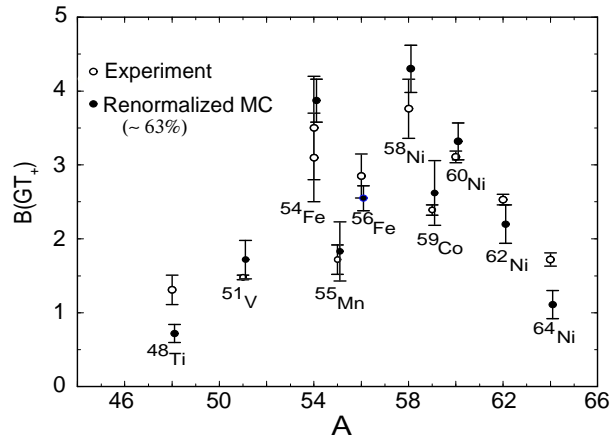}}
\caption{ Total Gamow-Teller strength $B(GT^+)$ in $pf$-shell nuclei. The 
solid  symbols are the renormalized SMMC strengths and the open symbols are 
the  experimental results.~\protect\cite{GT-exp} After Ref. 23. 
\label{fig:quenching}} 
\end{figure}

Calculations that include up to $2p-2h$ correlations account for some of the 
observed  quenching.\cite{ABB93}  To estimate how much of the quenching can be 
attributed  to proton-neutron correlations in the ground state, we performed 
SMMC  calculations in the
complete $pf$-shell for nuclei in the iron region.\cite{LD95}  The total 
strength  is renormalized by $(1/1.26)^2 \sim 63\%$ to account for the 
normalization  of the axial coupling constant from its free value of $g_A = 
1.26$  to  $g_A = 1$ in the nuclear medium. The renormalized strengths are 
shown  in Fig. \ref{fig:quenching} in comparison with the experimental values. 
We  see good agreement across the shell. We remark that a similar 
renormalization  of the Gamow-Teller operator in complete $sd$-shell 
calculations  also leads to an agreement with the data.\cite{BW}

  We conclude that any observed quenching 
 beyond an overall renormalization of $\sim 63 \%$ is due to proton-neutron 
correlations.  This has been confirmed in conventional shell model 
calculations  in the $pf$-shell, where the Gamow-Teller strength is found to 
decrease  as a function of the number of particle-hole excitations that are 
included  in the model space.\cite{CPZ96} 

  In SMMC it is also possible to calculate imaginary-time response functions 
and  invert them by the maxent method to obtain strength 
functions.\cite{LJK93}  Such methods were used to calculate Gamow-Teller 
strength  distributions for $pf$-shell nuclei.\cite{RDK97}

\section{$\gamma$-soft nuclei}\label{soft}

\begin{figure}[h!]
\hspace*{1 cm}\parbox{7. cm}{\epsfxsize=7. cm \epsfbox{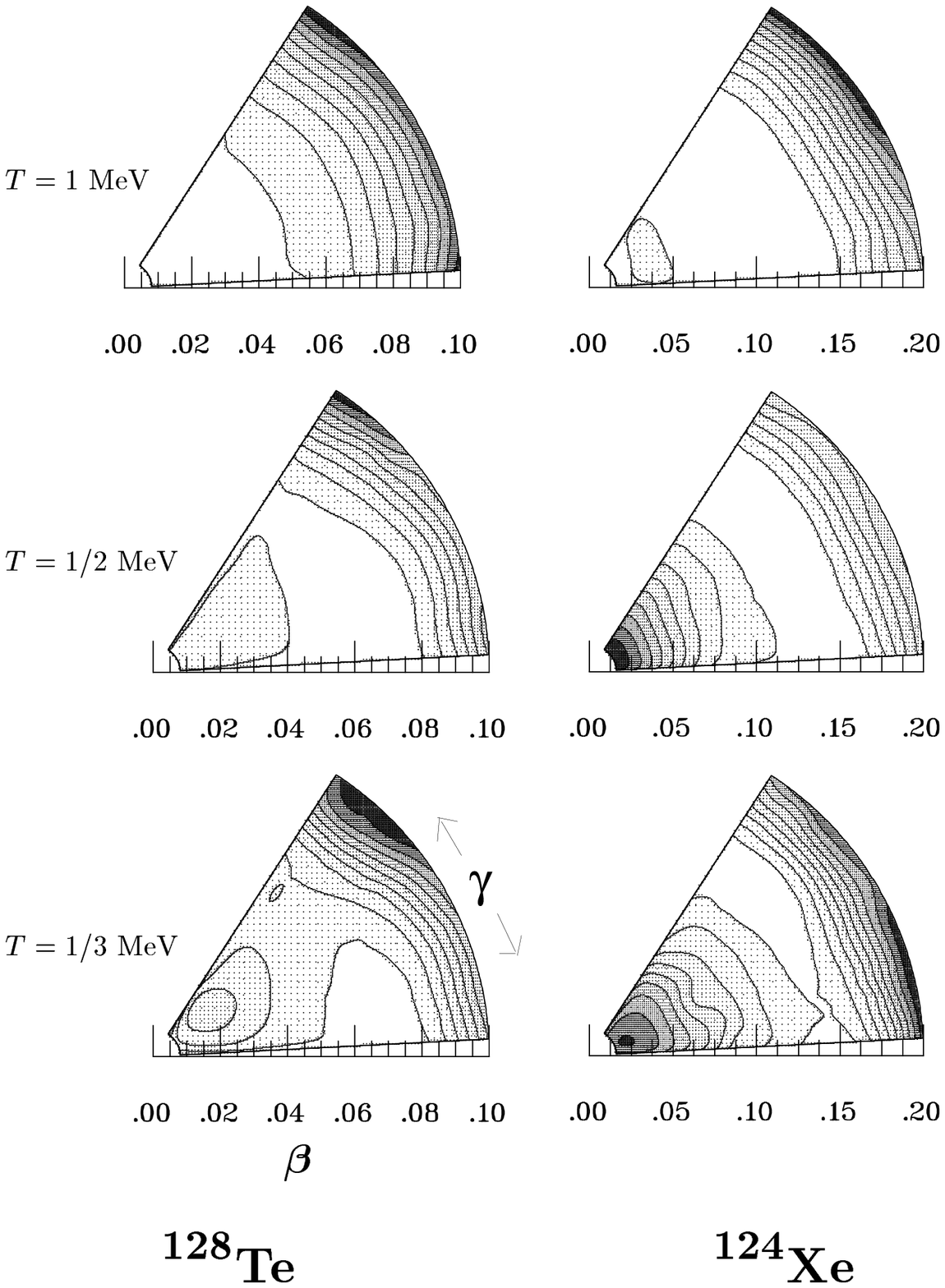}}
\parbox{4 cm}{\epsfxsize=4 cm \epsfbox{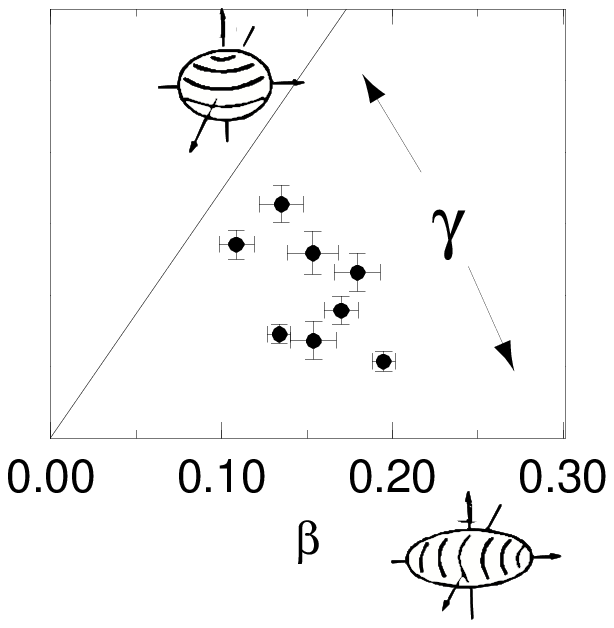}}
\caption{Free energy surfaces of $^{124}$Xe and $^{128}$Te at several 
temperatures.  The surfaces are inferred from shape distributions calculated 
in  SMMC. The inset on the right shows typical intrinsic shapes sampled in the 
Monte  Carlo together with their standard deviation. After Ref.~15. 
\label{fig:soft}} 
\end{figure}

One of the outstanding problems in nuclear structure is  to demonstrate that
  various  types of collectivity (known from phenomenological models
 of nuclear structure)  emerge in microscopic calculations.
 In particular, nuclei in the 100 -- 140 mass region are known to be 
$\gamma$-soft,  i.e., their potential energy is insensitive with respect to 
non-axial  deformations.  This region corresponds to the 50 -- 82 shell (for 
both  protons and neutrons), and could not be solved previously
 in fully microscopic calculations because of the prohibitively
  large size of the model space.  Using SMMC we have provided the first  
microscopic  evidence of softness
in this mass region\cite{ABDK}.  We have used the good-sign Hamiltonian 
(\ref{Hamiltonian})  and, among the various multipole interactions, kept only 
an  attractive quadrupole interaction  renormalized to account for core 
polarization.  A  quadrupole pairing interaction (in addition to monopole 
pairing)  was necessary to obtain correctly the excitation energy of the first 
$2^+$  state in tin isotopes.

To determine the quadrupole shape distribution $P(q_\mu)$, we calculated for 
each  sample $\sigma$ the average $\langle Q_\mu\rangle_\sigma$ and variance 
$\Delta_{\sigma}^2  =
\langle Q^2 \rangle_\sigma -\langle Q\rangle_\sigma^2 $ of the quadrupole 
mass  operator
$Q_\mu \equiv \sum r^2 {\cal Y}_{2\mu}$.
It is convenient to convert the shape distribution in the intrinsic frame 
$P(q^\prime_0,q^\prime_2)$   to an effective 
 free energy surface at finite temperature  using  the relation\cite{Al92}
 $P(q^\prime_0,q^\prime_2) dq^\prime_0 dq^\prime_2 \propto
\exp\left[- F(\beta, \gamma;T)/T\right] \beta^4 |\sin(3 \gamma)|
d \beta d \gamma$, where $\beta$ and $\gamma$ are the intrinsic  quadrupole 
shape  parameters.  Such free energy surfaces are shown in Fig. \ref{fig:soft} 
for  ${}^{128}$Te and ${}^{124}$Xe at several temperatures. The inset shows 
typical  intrinsic shapes,  demonstrating the weak dependence of the free 
energy   on $\gamma$.

\section{Level densities}\label{densities}

Level densities are important for theoretical estimates of nuclear reaction 
rates.  In particular, level densities are necessary input for nucleosynthesis 
calculations.  The $s$ and $r$ processes  are determined by the competition 
between  neutron capture and beta decay,\cite{BBF57} and the neutron-capture 
rates  are proportional to the level density of the corresponding compound 
nucleus.\cite{astro} 

Experimental data on level densities are available from different 
methods:\cite{Dilg}   direct counting of levels at low energies, neutron and 
proton  resonance data and charged particle spectra at intermediate energies, 
and  Ericson's fluctuations at higher energies.

 Most conventional calculations of level densities are based on Fermi gas 
models  in which important correlations are neglected. It is often found 
however,  that empirical modifications of the Fermi gas formula (also known as 
Bethe's  formula\cite{Bethe36}) can describe well experimental data. 
Particularly  useful is the backshifted Bethe formula (BBF) where the ground 
state  energy is backshifted by an amount $\Delta$. The backshift parameter 
originates  in pairing correlations and shell effects. A modified version of 
the  BBF is due to Lang and Le Couteur\cite{LLC54}
\begin{eqnarray}\label{BBF}
\rho (E_x) \approx
g  {{\sqrt\pi}\over{24}} a^{-\frac{1}{4}} (E_x - \Delta +t)^{-\frac{5}{4}}
e^{2\sqrt{a (E_x - \Delta)}} \;,
\end{eqnarray}
where $t$ is a thermodynamic temperature defined by
$E_x - \Delta = a t^2 - t$ and $g=2$.  The experimental level densities for 
many  nuclei are well described by (\ref{BBF}) when both $a$ and $\Delta$ are 
adjusted  for each nucleus. 
  It is therefore difficult to predict the level density to an accuracy 
better  than an order of magnitude.

  The nuclear shell model provides a suitable framework for the microscopic 
calculation  of level densities. The dimensionality of the model space is 
often  too large to allow for exact diagonalization, and truncations that 
might  be appropriate for low-lying states are not suitable at finite 
excitation  energies. Instead we  use the Monte Carlo methods described in 
Section  \ref{methods}.

  The calculations of level densities in SMMC\cite{NA97} are based on a 
thermodynamic  approach. The level density is obtained from the  canonical 
partition  function by an inverse Laplace transform
\begin{equation}\label{laplace}
\rho(E) = \int_{-i\infty}^{i\infty} \frac{d \beta}{2 \pi i} e^{\beta E}  
Z(\beta)  \;.
\end{equation}
 In practice we are interested in the average level density. It can be 
obtained  from (\ref{laplace}) in the saddle point approximation
\begin{eqnarray}\label{saddle}
\rho(E) \approx (2\pi \beta^{-2} C)^{-1/2}  e^{S}  \;,
\end{eqnarray}
where the canonical entropy $S$ and  heat capacity  $C$ are given by
\begin{equation}\label{thermo}
S = \ln Z + \beta E \;;\;\;\; C = \beta^2 (\langle H^2\rangle - \langle 
H\rangle^2)  =
  -\beta^2 \partial  E /\partial \beta \;.
\end{equation}

 In SMMC we calculate
the thermal energy $E(\beta)$ as an observable ($E(\beta) \equiv\langle 
H\rangle_\beta$),  and then find the partition function by integrating the 
exact  thermodynamic relation $E = -\partial \ln Z/\partial \beta$:
 \begin{eqnarray}
 \ln \left[ Z(\beta)/ Z(0)\right] =
 - \int_0^\beta d\beta^\prime E(\beta^\prime) \;.
\end{eqnarray}
 $Z(0) = {\rm Tr }~ {\bf 1}$ is the total number of states in the model 
space. 
  The entropy and heat capacity are then calculated from Eqs. (\ref{thermo}) 
to   give the level density (\ref{saddle}).

\subsection{Level densities in the iron region}

 We have used the methods of Section \ref{densities} to calculate  level 
densities  for nuclei in the iron region. The model space includes the 
$pf+g_{9/2}$-shell,  and is good for excitation energies up to $E_x \sim 20$ 
MeV.  With the inclusion of the $g_{9/2}$ orbit we can calculate both even and 
odd  parity states.

 We used the good-sign Hamiltonian (\ref{Hamiltonian}) and kept only the 
quadrupole,  octupole and hexadecupole terms with renormalization factors of 
$k_2=2$,  $k_3=1.5$ and $k_4=1$.
 Using experimental odd-even mass differences we determined
 $g_0 \approx 0.2$ MeV. 

\begin{figure}[h!]
%\figurebox{22pc}{15pc}{}
\epsfxsize=8 cm
\centerline{\epsfbox{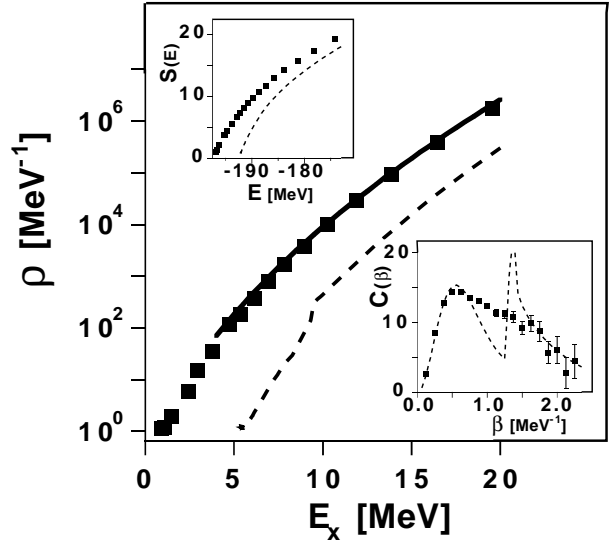}}
\caption{Total level density of $^{56}$Fe versus excitation energy $E_x$: 
SMMC  (solid squares), experiment (solid line) and HF (dashed line). The top 
inset  shows the entropy $S(E)$,  and the bottom inset is the heat capacity 
$C(\beta)$.  The solid squares are the SMMC results while the dashed lines are 
obtained  in the HF approximation. After Ref.~16. \label{fig:fe56}}
\end{figure}

Fig. \ref{fig:fe56} demonstrates the calculation of the level density of 
$^{56}$Fe.  The top and bottom insets show the entropy versus energy and the 
heat  capacity versus $\beta$, respectively. For comparison we show the 
Hartree-Fock  (HF) results by dashed lines. In the HF we observe a 
discontinuity  in the heat capacity around $\beta \sim 1.3$ MeV$^{-1}$ due to 
a  shape transition from a deformed to spherical nucleus. This discontinuity 
is  washed out by the fluctuations in the SMMC calculations. The total SMMC 
level  density for $^{56}$Fe is shown in Fig. \ref{fig:fe56} versus excitation 
energy  $E_x$ (solid squares). The solid line is inferred from the 
experiment.\cite{LVH72}  We see good agreement between the microscopic 
calculations  and the data, without any adjustable parameters.

\begin{figure}[h]
%\figurebox{22pc}{15pc}{}
\epsfxsize= 12 cm
\centerline{\epsfbox{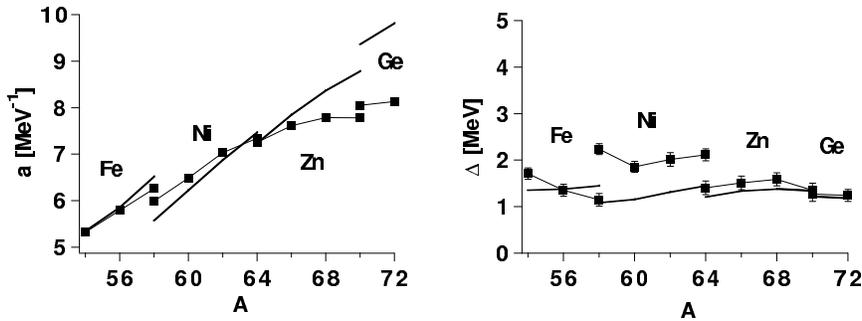}}
\caption{Single-particle level density parameter $a$ (left) and backshift 
parameter  $\Delta$ (right) as a function of mass number $A$, obtained by 
fitting  the SMMC level densities to the BBF. The solid lines are the 
empirical  values of Ref.~34. Shown are results for even-even nuclei from Fe 
to  Ge. After Ref.~33. \label{fig:a-del}}
\end{figure}

  The SMMC level density can be well-fitted to the BBF. We can therefore 
extract  $a$ and $\Delta$ from the microscopic calculations. Fig. 
\ref{fig:a-del}  shows the parameters $a$ (left) and $\Delta$ (right) as a 
function  of mass number for even-even nuclei in the mass range $A \sim 50 - 
70$.\cite{NA98}  The SMMC results (solid squares) are compared with the 
empirical  values of Ref.~34 (solid lines). We observe that $a$ varies 
smoothly  with mass, but that $\Delta$ shows shell effects and is enhanced at 
$N=28$  or $Z=28$.

\subsection{Level densities by particle-number reprojection}

 The projection on an odd number of particles introduces a new sign problem, 
even  for good-sign Hamiltonians. Moreover, the calculations of level 
densities  are time-consuming since each nucleus requires new Monte Carlo 
calculations  for all temperatures.

The particle-number reprojection method of Section \ref{reproject} can be 
used  to carry out the Monte Carlo sampling for an even-even (or $N=Z$) 
nucleus  ${\cal A}$ followed by a reprojection to find the thermal energies 
for  a series of nuclei ${\cal A'}$. We have applied this method to calculate 
the  level densities of manganese, iron and cobalt isotopes by reprojecting 
from  $^{56}$Fe and $^{54}$Co.\cite{ALN99}   Since the Hamiltonian 
(\ref{Hamiltonian})  depends on ${\cal A}$ (mostly through its single-particle 
spectrum),  suitable corrections should be made to the thermal energy.

\begin{figure}[h]
%\figurebox{22pc}{15pc}{}
\epsfxsize= 9 cm
\centerline{\epsfbox{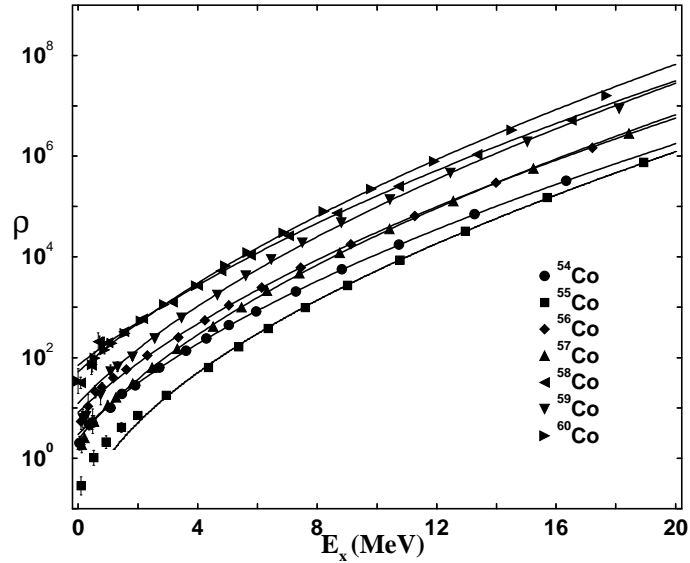}}
\caption{Level densities of cobalt isotopes versus excitation energies 
obtained  in the particle-number reprojection method. The solid lines are fits 
to  the BBF formula (\protect\ref{BBF}). After Ref.~19.\label{fig:Co-level}}
\end{figure}
\begin{figure}[h]
%\figurebox{22pc}{15pc}{}
\epsfxsize= 10 cm
\centerline{\epsfbox{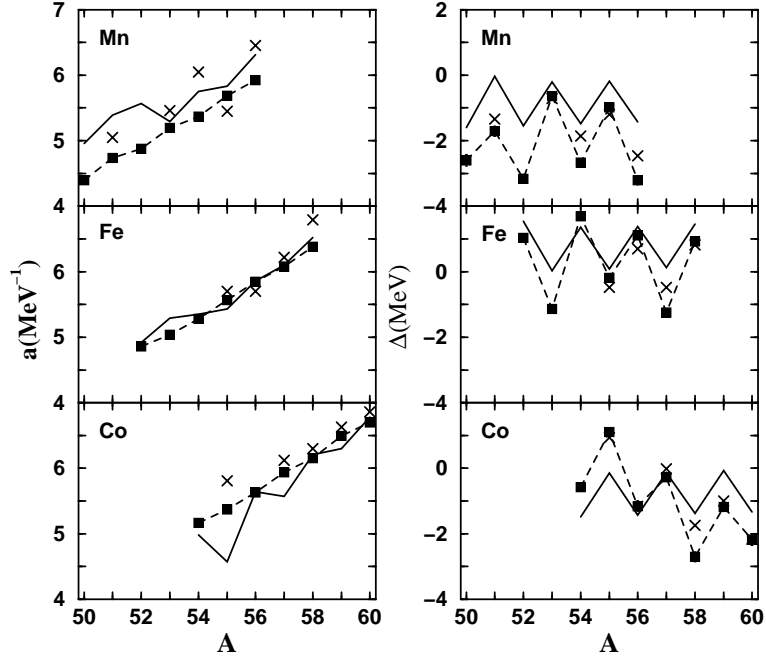}}
\caption{The single-particle level density parameter $a$ (left) and backshift 
parameter  $\Delta$ (right) in manganese, iron and cobalt isotopes. The values 
determined  by particle-number reprojection SMMC  (solid squares) are compared 
with  the experimental values (x's)~\protect\cite{Dilg} and the empirical 
values  of Ref.~34 (solid lines). After Ref.~19.\label{fig:a-delta}}
\end{figure}

Results for the level densities of cobalt isotopes, reprojected from 
$^{56}$Fe,  are shown in Fig. \ref{fig:Co-level}.  We observe that the level 
density  of an odd-odd cobalt (e.g., $^{54}$Co) is higher than the level 
density  of the subsequent even-odd cobalt ($^{55}$Co) although the latter has 
a  larger mass.
The systematics of $a$ and $\Delta$ is shown in Fig. \ref{fig:a-delta} for 
manganese,  iron and cobalt isotopes, including odd-$A$ and odd-odd nuclei. 
The  SMMC results (solid squares) are compared with the experimental values 
(x's)  and the empirical formulas of Ref.~34 (solid lines). The staggering 
effect  in $\Delta$ originates in pairing correlations. The microscopic 
calculations  are often in better agreement with the data than are the 
empirical  values.

\subsection{Parity-projected level densities}

We have calculated the parity dependence of level densities  by introducing 
parity  projection in the HS representation.  The projection operators on 
even-  and odd-parity states are given by $P_\pm = (1\pm P)/2$, where $P$ is  
the  parity operator. The parity-projected thermal
energies $E_\pm (\beta) \equiv { {\rm Tr} (H P_\pm e^{-\beta H} )
 / {\rm Tr} (P_\pm e^{-\beta H})}$ can be written as
\begin{eqnarray} \label{E_pm}
 E_\pm (\beta)  = {\int D[\sigma]W_\sigma \Phi_\sigma  \left[ \langle H 
\rangle_\sigma  \pm
\langle H \rangle_{P\sigma} \zeta_P(\sigma) / \zeta(\sigma) \right]
\over
\int D[\sigma]W_\sigma \Phi_\sigma \left[  1 \pm {\zeta_P(\sigma) /
\zeta(\sigma)} \right] }
\;,
\end{eqnarray}
where $\zeta(\sigma) \equiv {\rm Tr}\; U_\sigma$, $\zeta_P(\sigma) \equiv 
{\rm 
Tr}\;(PU_\sigma)$ and
$\langle H \rangle_{P\sigma} \equiv {\rm Tr}\;(HPU_\sigma)/{\rm 
Tr}\;(PU_\sigma)$. 
 $PU_\sigma$ can be represented
in the single-particle space by the matrix ${\bf P} {\bf U}_\sigma$, where 
${\bf  P}$ is  a diagonal matrix with elements $(-)^{\ell_a}$
($\ell_a$ is the orbital angular momentum of the single-particle orbit $a$). 
Consequently,  the integrand in (\ref{E_pm}) can be calculated by matrix
algebra in the single-particle space. From $E_\pm(\beta)$ we  calculate the 
positive-  and negative-parity level densities using the method discussed in 
Section  \ref{densities}.

\begin{figure}[h]
\epsfxsize= 11 cm
\centerline{\epsfbox{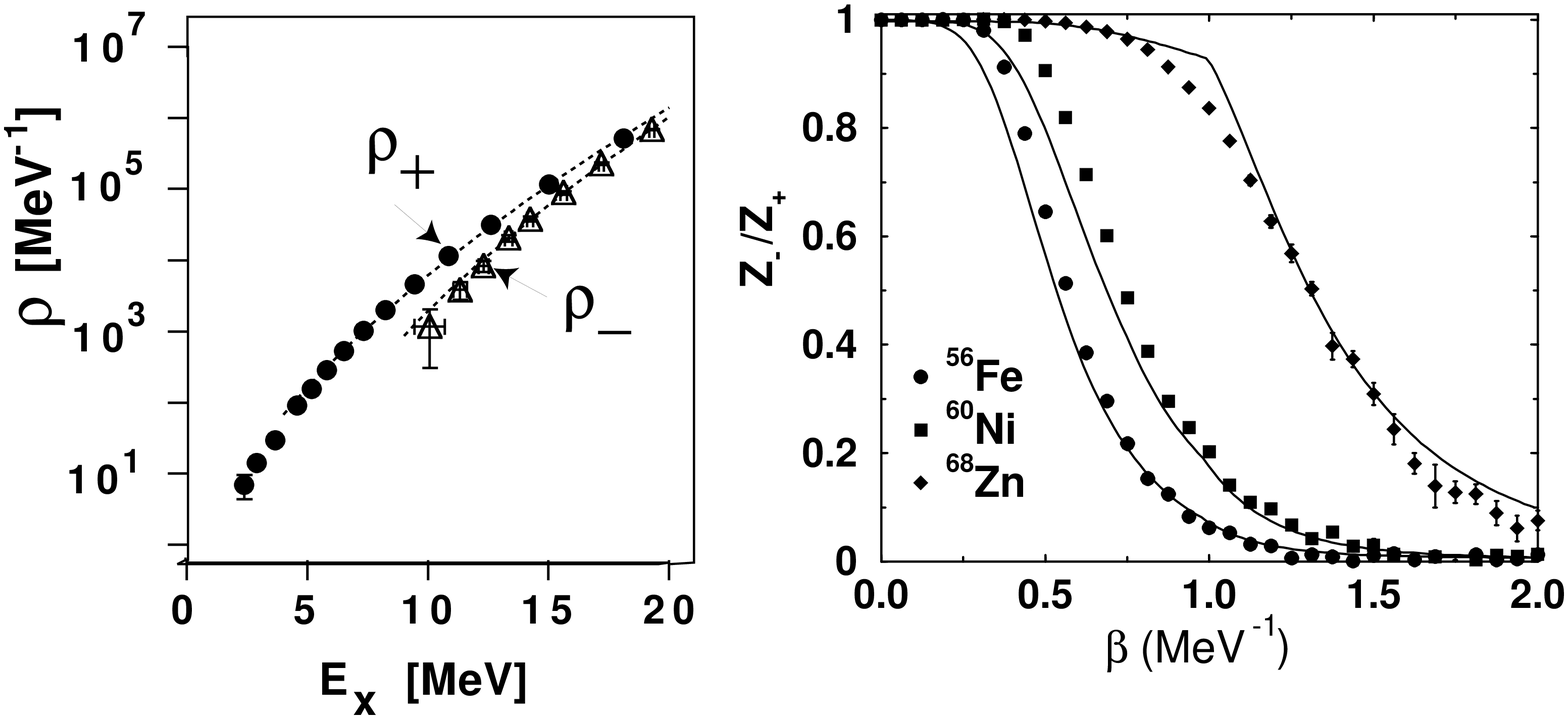}}
\caption{ Left: parity projected level densities $\rho_\pm$ in $^{56}$Fe. 
Right:  odd-to-even partition function ratio versus $\beta$ for $^{56}$Fe, 
$^{60}$Ni  and $^{68}$Zn. The symbols are the SMMC calculations and the solid 
lines  are obtained from Eq. (\protect\ref{parity-ratio}) of the statistical 
quasi-particle  model (see text). After Refs.~16 and 35.\label{fig:ratio-MC}}
\end{figure}

  The left panel of Fig. \ref{fig:ratio-MC} shows the parity-projected level 
densities  for $^{56}$Fe. We see that $\rho_+\neq \rho_-$ even at the neutron 
resonance  energies. This is contrary to the common assumption of equal even 
and  odd parity densities, often used in nucleosynthesis calculations.

It is difficult to calculate the odd-parity level density for the even-even 
nuclei   at low excitation energies because of a new sign problem introduced 
by  the projection $P_-$. The sign problem affects less the odd-to-even 
partition  function ratio.  This ratio is calculated from
\begin{equation}\label{ratio-MC}
{Z_-\over Z_+} = \left.\left[ 1 - \left\langle
{\zeta_P(\sigma) \over \zeta(\sigma)} \right\rangle_W \right]
\right/\left[ 1 + \left\langle
{\zeta_P(\sigma) \over \zeta(\sigma)} \right\rangle_W \right]\;,
\end{equation}
and shown in the right panel of Fig. \ref{fig:ratio-MC} for several nuclei in 
the  iron region.

The parity distribution as a function of temperature (or excitation energy) 
can  be understood in terms of a simple statistical  model.\cite{parity}
The single-particle states are divided into two groups of odd- and 
even-parity  states.
Assuming that the particles occupy the single-particle states randomly and 
independently,  the probability to find $n$ nucleons in states
with parity $\pi$ ($\pi$ is the parity of the group with the smaller 
occupation)   is a Poisson distribution,
$P(n) = f ^{n} e^{-f}/ n!$, where $f=\langle n\rangle$.
For an even-$A$ nucleus, an even (odd) parity many-particle state is obtained 
when  $n$ is even (odd).  We find a simple formula for the odd-to-even parity 
ratio 
\begin{equation}\label{parity-ratio}
{Z_-(\beta) \over Z_+(\beta)} = {\sum_{n \;\rm odd} P(n) \over \sum_{n \;\rm 
even}  P(n)} =\tanh f \;.
\end{equation}
$f$ is estimated from the Fermi-Dirac occupations of deformed single-particle 
states  with parity $\pi$. The distribution $P(n)$ can be calculated in SMMC 
by  projection on states with parity $\pi$\cite{parity}. In the iron region we 
find  good agreement with the Poisson distribution above the pairing 
transition  temperature of $\sim 1$ MeV. Below this temperature, our model is 
still  valid but for the quasi-particles. $f$ is now given by\cite{BCS}
 $f =  \sum_{a \in \pi} f_a = \sum_{a \in \pi} [1+ e^{\beta E_a}]^{-1}$ where
$E_a=\sqrt{(\epsilon_a-\lambda)^2 +\Delta^2}$ are the quasi-particle energies 
(and  $\lambda$ is the chemical potential). The solid lines in Fig. 
\ref{fig:ratio-MC}  are the result (\ref{parity-ratio}) of the quasi-particle 
model. 

\section*{Conclusions}

 We have discussed a quantum Monte Carlo approach to the interacting shell 
model  at finite and zero temperature. The initial applications of these 
techniques  have been severely limited by the Monte Carlo sign problem. A 
practical  solution to the sign problem in the nuclear case has allowed new 
realistic  calculations that could not be carried out in conventional 
diagonalization  methods.  Alternatively, good-sign interactions can be 
constructed  for realistic estimates of certain nuclear properties (e.g., 
collective  properties and level densities).
We have presented a variety of applications:
the quenching of the Gamow-Teller strength,
microscopic evidence of $\gamma$-softness, and accurate
calculations of level densities and their parity distribution.

Finally, it is important to keep in mind that the Monte Carlo approach is 
complementary  to the conventional shell model approach; it cannot be used to 
calculate  detailed spectra but on the other hand it is very useful for finite 
and  zero temperature calculations. 

\section*{Acknowledgments}

This work was supported in part by the Department of Energy grant
No.\ DE-FG-0291-ER-40608.
I  would like to thank  G.F. Bertsch,  D.J. Dean,  S.E. Koonin, K. Langanke, 
S.  Liu, H. Nakada and W.E. Ormand for their collaboration on various parts of 
the  work presented above.

\end{document}